\documentclass[journal]{IEEEtran}

\usepackage{cite}
\usepackage[caption=false,font=footnotesize]{subfig}
\usepackage{graphicx}
\usepackage[cmex10]{amsmath}
\usepackage{amsfonts,amssymb,amsthm}
\usepackage{mathrsfs}
\usepackage{enumerate}
\usepackage[makeroom]{cancel}
\usepackage{color}

\usepackage[utf8]{inputenc}
\usepackage[english]{babel}
\usepackage[percentage]{overpic}

\usepackage{tikz}
\usepackage{verbatim}
\usetikzlibrary{shapes.gates.logic.US,trees,positioning,arrows}

\newtheorem{proposition}{Proposition}

\theoremstyle{remark}
\newtheorem*{remark}{Remark}

\begin{document}

\title{Rethinking~the~Intercept~Probability\\of Random Linear Network Coding}

% -- START: Deactivate for submission / Re-activate for camera-ready
\author{Amjad~Saeed~Khan, Andrea~Tassi and Ioannis~Chatzigeorgiou% <-this % stops a space

\thanks{This work was carried out under the auspices of COST Action IC1104 and the support of EPSRC under Grant EP/L006251/1. 

A. S. Khan and I. Chatzigeorgiou are with the School of Computing and Communications, Lancaster University, LA1~4WA, UK (e-mail: \{a.khan9, i.chatzigeorgiou\}@lancaster.ac.uk).

A. Tassi is with the Department of Electrical and Electronic Engineering, University of Bristol, BS8~1UB, UK (email: a.tassi@bristol.ac.uk).}}% <-this % stops a space
% -- STOP -- 

\maketitle

\begin{abstract}
This letter considers a network comprising a transmitter, which employs random linear network coding to encode a message, a legitimate receiver, which can recover the message if it gathers a sufficient number of linearly independent coded packets, and an eavesdropper. Closed-form expressions for the probability of the eavesdropper intercepting enough coded packets to recover the message are derived. Transmission with and without feedback is studied. Furthermore, an optimization model that minimizes the intercept probability under delay and reliability constraints is presented. Results validate the proposed analysis and quantify the secrecy gain offered by a feedback link from the legitimate receiver.
\end{abstract}


% -----------------------------------------------------------------------------

\section{Introduction}
\label{sec:intro}

In the context of networks and protocols, network coding~\cite{Ahlswede2000} has been widely recognized as an intriguing technique to improve network performance. It can considerably reduce transmission delay, processing complexity and energy consumption, and has the potential to significantly increase throughput and robustness~\cite{Bassoli2013}. Therefore, it has been studied for use in many applications, including large scale content distribution in peer-to-peer networks~\cite{Gkantsidis2005} and data transmission in sensor networks or delay tolerant networks~\cite{Katti2007}. Due to the broadcast nature of wireless channels, networks are vulnerable to security attacks, such as wiretapping and eavesdropping. The problem of achieving secure communication in systems employing network coding has recently attracted the attention of the research community in wireles networks. Ning and Yeung~\cite{Cai2002} first formulated the concept of secure network coding, which avoids information leakage to a wiretapper. They imposed a security requirement, that is, the mutual information between the source symbols and the symbols received by the wiretapper must be zero for secure communication. Based on a well-designed precoding matrix, Wang~\textit{et al.}~\cite{Wang2013} proposed a secure broadcasting scheme with network coding to obtain perfect secrecy. Probabilistic weak security for linear network coding was presented in~\cite{Adeli2013}, which devised network coding rules that can improve security depending on the adopted field size, the number of transmitted symbols and the ability of the attacker to eavesdrop on one or more independent channels.

Recently, the intercept probability of fountain coding, which is equivalent to random linear network coding for wireless broadcast applications, was formulated in~\cite{Niu2014}. Our work has been inspired by the methodology in~\cite{Niu2014} but differs in two major points. Firstly, we have revisited the derivation of the intercept probability. More specifically, the decoding probability of a receiver has been taken into account in our calculations. Furthermore, key probability expressions have been revised to accurately reflect (i) the effect of the size of the finite field over which network coding is performed, (ii) the impact of a feedback link between the legitimate receiver and the transmitter, and (iii) the fact that the number of transmitted coded packets cannot be infinite in practice. The second difference is that~\cite{Niu2014} proposed an optimization model with respect to the number of source packets composing a message. However, the number of source packets and, by extension, their length are often dictated by the provided service. Our objective is to minimize the intercept probability by optimizing the number of transmitted coded packets, under delay and reliability constraints. As part of the optimization process, we prove that awareness of the existence of an eavesdropper is not required by the transmitter and the legitimate receiver.

%The rest of the letter is structured as follows. Section~\ref{sec:system} describes the system model. Section~\ref{sec:analysis} derives theoretical expressions for the intercept probability of the network. Section~\ref{sec:optimization} formulates and solves the proposed resource allocation problem. Numerical results are presented and discussed in Section~\ref{sec:results}, and conclusions are drawn in Section~\ref{sec:conclusion}. 

% -----------------------------------------------------------------------------

\section{System Model}
\label{sec:system}

% -- MOVE Fig.1 HERE for submission

\begin{figure}[t]
\centering
\includegraphics[width=0.6\columnwidth]{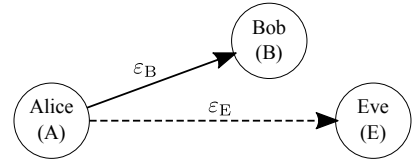}
\caption{Block diagram of the system model, where $\varepsilon_\textrm{B}$ and $\varepsilon_\textrm{E}$ denote the erasure probabilities of the channels linking Alice to Bob and Alice to Eve, respectively.}
%\vspace{-4mm}
\label{fig:fig_sys}
\end{figure}

We consider a network configuration whereby a source (Alice) wishes to transmit a message to a legitimate destination (Bob) in the presence of a passive eavesdropper (Eve), as shown in Fig. 1. Before initiating the communication process, Alice segments the message into $K$ source packets and employs Random Linear Network Coding (RLNC) to generate and broadcast $N \geq K$ coded packets. The links connecting Alice to Bob and Alice to Eve are modeled as packet erasure channels characterized by erasure probabilities $\varepsilon_\textrm{B}$ and $\varepsilon_\textrm{E}$, respectively. As per the RLNC requirements, Bob and Eve can recover the message only if they collect at least $K$ linearly independent coded packets. 

Based on this setup and the general condition that $\varepsilon_\textrm{B}< \varepsilon_\textrm{E}$ for physical layer security, we consider two network coded transmission modes, which we refer to as \emph{Feedback-aided Transmission} (FT) and \emph{Unaided Transmission} (UT). In the FT mode, Alice broadcasts up to $N$ coded packets but ceases transmission as soon as Bob sends a notification over a perfect feedback channel acknowledging receipt of $K$ linearly independent coded packets. In the case of UT, a feedback channel between Bob and Alice is not available, therefore Alice broadcasts exactly $N$ coded packets anticipating Bob to successfully recover her message. In both modes, the communication process is considered to be secure if Eve fails to reconstruct Alice's message. In the rest of this letter, we will investigate the resilience of FT and UT to the interception of $K$ linearly independent coded packets by Eve.

% -- MOVE Fig.1 HERE for camera-ready

% -----------------------------------------------------------------------------

\section{Performance Analysis}
\label{sec:analysis}

The physical layer security offered by the two transmission modes will be quantified by the probability that Eve will manage to recover the message. To derive this probability, which is known as the \emph{secrecy outage probability} or the \emph{intercept probability}, we will first consider the general case of point-to-point communication between Alice and a receiver $D$ over an erasure channel with erasure probability $\varepsilon_D$. Note that $D$ can be either Bob or Eve, i.e., $D\in\{\textrm{B},\textrm{E}\}$. If Alice transmits $N\geq K$ coded packets and the receiver retrieves $n_\textrm{R}$ coded packets, where $K\leq n_\textrm{R}\leq N$, the probability that the receiver will successfully recover the $K$ source packets is given by \cite{Trullols-Cruces2011}
\begin{equation}
\label{eq:decoding_prob}
P(n_\textrm{R},K) = \prod_{i=0}^{K-1}\left[1-q^{-\left(n_\textrm{R}-i\right)}\right],
\end{equation}
where $q$ is the size of the finite field over which network coding operations are performed. Let $X$ be a random variable that represents the number of transmitted coded packets for which the receiver can recover the $K$ source packets. The Cumulative Distribution Function (CDF) of $X$ describes the probability that the receiver will recover the $K$ source packets after $n_\textrm{T}$ coded packets have been transmitted, where $K\leq n_\textrm{T}\leq\!N$. This CDF can be obtained by averaging \eqref{eq:decoding_prob} over all valid values of $n_\textrm{R}$, that is,
\begin{equation}
\label{eq:prob_cdf}
\begin{split}
F_D(n_\textrm{T})&=\textrm{Pr}\left\{X \leq n_\textrm{T}\right\}\\ 
&=\!\sum_{n_\textrm{R}=K}^{n_\textrm{T}}\!\binom{n_\textrm{T}}{n_\textrm{R}}(1-\varepsilon_D)^{n_\textrm{R}}\varepsilon_D^{n_\textrm{T}-n_\textrm{R}}\;P(n_\textrm{R},K).
\end{split}
\end{equation}
The probability that the receiver will recover the $K$ source packets when the $n_\textrm{T}$-th coded packet has been transmitted, but not earlier, is given by the Probability Mass Function (PMF) of $X$, which can be derived as follows:
\begin{equation}
\begin{split}
\label{eq:prob_pmf}
f_D(n_\textrm{T})&= \textrm{Pr}\left\{X = n_\textrm{T}\right\}\\ 
&=\left\{
	\begin{array}{ll}
		\!F_D(n_\textrm{T})-F_D(n_\textrm{T}-1), &\!\!\!\mbox{if }K<n_\textrm{T}\leq N\\[0.5em]
		\!F_D(K), &\!\!\!\mbox{if }n_\textrm{T} = K.
	\end{array}
\right.
\end{split}
\end{equation}

Let us now return our focus to the considered network configuration operating in the FT mode. Recall that Bob sends an acknowledgement to Alice when he receives $K$ linearly independent coded packets and can thus recover the source message. The intercept probability can be expressed as the sum of two constituent probabilities:
\begin{equation}
\label{eq:prob_ic_FT}
P^{\,\textrm{FT}}_\textrm{int}(N) = P_{\,\textrm{BE}}(N) + P_{\,\textrm{E}}(N).
\end{equation}
The first term of the sum in \eqref{eq:prob_ic_FT}, $P_{\,\textrm{BE}}(N)$, denotes the probability that both Bob and Eve will recover the message. This can happen if Bob decodes the message only after the \mbox{$n_\textrm{T}$-th} coded packet has been transmitted, while Eve has already recovered the message or recovers it concurrently with Bob. Invoking the definitions in \eqref{eq:prob_cdf} and \eqref{eq:prob_pmf}, and considering all possible values of $n_\textrm{T}$, we can express $P_{\,\textrm{BE}}(N)$ as
\begin{equation}
\label{eq:prob_BE}
P_{\,\textrm{BE}}(N)=\sum_{n_\textrm{T}=K}^{N}f_\textrm{B}(n_\textrm{T})\,F_\textrm{E}(n_\textrm{T}).
\end{equation}
The second term of the sum in \eqref{eq:prob_ic_FT}, $P_{\,\textrm{E}}(N)$, represents the probability that Eve will be successful in recovering the message but Bob will fail to decode it after Alice has transmitted the complete sequence of $N$ coded packets. Using the CDF of the number of coded packets delivered by Alice to Eve and Bob, respectively, we can write $P_{\,\textrm{E}}(N)$ as follows:
\begin{equation}
\label{eq:prob_E}
P_{\,\textrm{E}}(N) = F_\textrm{E}(N)\,\left[\,1-F_\textrm{B}(N)\,\right].
\end{equation}
We should stress that \eqref{eq:prob_BE} and \eqref{eq:prob_E} are exact only if the sequence of coded packets delivered over the Alice-to-Bob link is independent of the sequence delivered over the Alice-to-Eve link. This is a common hypothesis in the literature of broadcast networks, e.g., \cite{Niu2014} and \cite{Kurniawan2010}, and is valid for a non-vanishing product between the number of coded packets transmitted over a channel and the erasure probability of that channel \cite{Khan2015}. The accuracy of \eqref{eq:prob_ic_FT} will also be demonstrated in Section \ref{sec:results}.

In the case of UT, a feedback channel is not available between Bob and Alice, therefore Alice transmits the complete sequence of $N$ coded packets uninterruptedly. Therefore, the intercept probability is simply equal to the probability that Eve will recover the message after Alice has transmitted $N$ coded packets. Using the definition of the CDF in \eqref{eq:prob_cdf}, we obtain
\begin{equation}
\label{eq:prob_ic_UT}
P^{\,\textrm{UT}}_\textrm{int}(N) = F_\textrm{E}(N).
\end{equation}

Manipulation of the expression for $P^{\,\textrm{FT}}_\textrm{int}(N)$, as shown in Appendix \ref{sec:rewrite_FT}, and subtraction of $P^{\,\textrm{UT}}_\textrm{int}(N)$ from it, yields
\begin{equation}
\label{eq:diff}
P^{\,\textrm{FT}}_\textrm{int}(N)-P^{\,\textrm{UT}}_\textrm{int}(N) =\, 
-\!\!\!\!\sum_{n_\textrm{T}=K+1}^{N}\!\!\!\!f_\textrm{E}(n_\textrm{T})\,F_\textrm{B}(n_\textrm{T}-1).
\end{equation}
Expression \eqref{eq:diff} measures the loss in the intercept capability of Eve or, equivalently, the gain in secrecy by Bob, if Bob can acknowledge the recovery of the source message to Alice using a feedback channel.

\begin{remark}
In this letter, we assume that Alice has knowledge of the \emph{average} channel conditions, characterized by the erasure probability, between her and Bob. If Alice could sense the \emph{instantaneous} channel quality and transmitted coded packets only when the channel quality warranted their \mbox{error-free} delivery to Bob, as in \cite{Niu2014}, \cite{Guan2015}, the equivalent erasure probability of the link between Alice and Bob would be $\varepsilon_\textrm{B}=0$. In that case, Alice could generate exactly $K$ linearly independent coded packets in a deterministic manner, as opposed to random, and forward them to Bob. As a result, the intercept probability would reduce to $(1-\varepsilon_\textrm{E})^{K}$ regardless the transmission mode. This remark concurs with the conclusion of \cite{Niu2014} that an arbitrarily small intercept probability can be achieved by increasing the value of $K$, but at the cost of increased delay.
\end{remark}

% -----------------------------------------------------------------------------

\section{Optimization Model}
\label{sec:optimization}

This section aims to determine the optimum value of $N$, i.e., the number of coded packet transmissions, that minimizes the intercept probability, provided that a hard deadline is met. This hard deadline, denoted by $\Hat{N}$, represents the number of coded packet transmissions that Alice is not allowed to exceed. In addition, the proposed optimization strategy permits Bob to recover the message with a target probability $\Hat{P}$. In the rest of this letter, both FT and UT will be optimized by the Resource Allocation Model (RAM), which is defined as follows:
\begin{align}
	\text{(RAM)} &  \quad  \min_{N} \,\,  P_\textrm{int}(N)\label{eq:objective}\\
    \text{subject to} &   \quad F_\textrm{B}(N) \geq \Hat{P}\label{eq:constraint_1}\\
                      &   \quad N \leq \Hat{N} \label{eq:constraint_2}
\end{align}
where the objective function~\eqref{eq:objective} represents the intercept probability when $N$ coded packets have been scheduled for transmission. Constraint~\eqref{eq:constraint_1} ensures that the probability of Bob recovering the message is at least $\Hat{P}$, while constraint~\eqref{eq:constraint_2} imposes that the number of planned coded packet transmissions is less than or equal to $\Hat{N}$. 
%Parameters that do not contribute to the optimization problem have been dropped from the notation but indices signifying the receiving node have been retained, e.g., $F(N,K,\varepsilon_\textrm{B})$ has been simplified to $F_\textrm{B}(N)$.

% -- MOVE Fig.2 HERE for submission

% --- START of FIG2 ---
\begin{figure}[t]
\centering
\includegraphics[width=1\columnwidth]{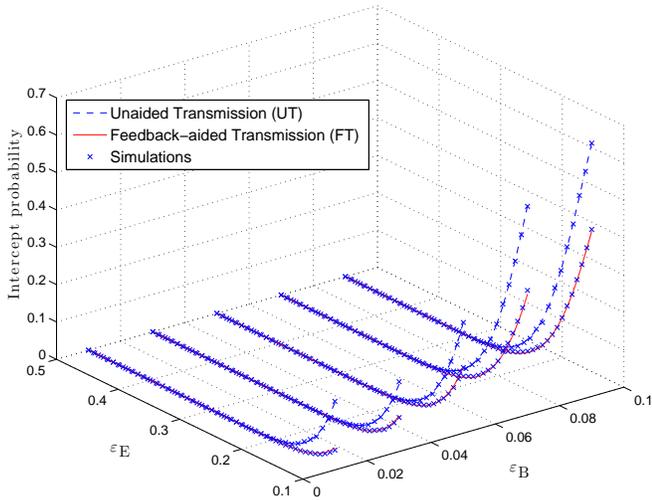}
\caption{Comparison between analytical and simulation results for FT and UT, when $\varepsilon_\textrm{E}\in[0.1,0.5]$, $\varepsilon_\textrm{B}= \{0.01,0.03,0.05,0.07,0.09\}$, $K=50$, $\hat{N}=150$, $q=2$ and $\hat{P}=90$.}
%\vspace{-5mm}
\label{fig:fig_compare}
\end{figure}
% --- END of FIG2 ---

The proof of the following proposition will contribute to the solution of the RAM problem.
\begin{proposition}
\label{pro:prop1}
The intercept probability $P_\textrm{int}(N)$ is a non-decreasing function of $N$, i.e., 
\begin{equation}
\label{eq:proposition}
P_\textrm{int}(N_1) \leq P_\textrm{int}(N_2)\quad\textrm{for all}\quad N_1\leq N_2.
\end{equation}
\end{proposition}

\begin{IEEEproof} 
One of the properties of CDFs is that they are non-decreasing functions and, as per \eqref{eq:prob_ic_UT}, the intercept probability of UT is equal to a CDF. In the case of FT, the subtraction of $P_\textrm{int}(N_1)$ from $P_\textrm{int}(N_2)$ for $N_2\geq N_1$ gives a sum of non-negative terms, as shown in Appendix \ref{sec:proofProp}. Therefore, \mbox{$P_\textrm{int}(N_2)-P_\textrm{int}(N_1)\!\geq\!0$}, which concludes the proof.
\end{IEEEproof}

\vspace{2mm} % Reduce to 1 mm for camera-ready

We can now proceed to Proposition~\ref{pro:prop2} and provide a description of the solution to the RAM problem.
\begin{proposition}
\label{pro:prop2}
If the RAM problem admits a solution, the optimum solution is
\begin{equation}
\label{eq:optimum_sol}
N^* = \arg\min \left\{N\in [K,\Hat{N}] \,\,\big|\,\, F_B(N) \geq \Hat{P}\right\}.
\end{equation}
\end{proposition}

\begin{IEEEproof} 
Let $N^*$ denote the smallest value of $N$ in the interval $[K,\hat{N}]$ for which constraint~\eqref{eq:constraint_1} holds. If an integer value smaller than $N^*$ is selected, for example $N^*-1$, the intercept probability will reduce, as per Proposition \ref{pro:prop1}, but constraint \eqref{eq:constraint_1} will not be met. We thus conclude that $N^*$ is the optimum solution to the RAM problem.
\end{IEEEproof} 

Root-finding algorithms, such as the bisection method, can be used on the right-hand side of \eqref{eq:optimum_sol} to determine if $N^*$ exists and identify its value. Based on this analysis, we showed that minimization of the intercept probability under delay and reliability constraints can be achieved by minimizing the number of transmitted coded packets. Thus, Alice should know the erasure probability of the channel between her and Bob but knowledge of the presence of an eavesdropper \emph{is not necessary}.

% -- MOVE Fig.3 HERE for submission

% --- START of FIG3 ---
\begin{figure}[t]
\centering
\includegraphics[width=1\columnwidth]{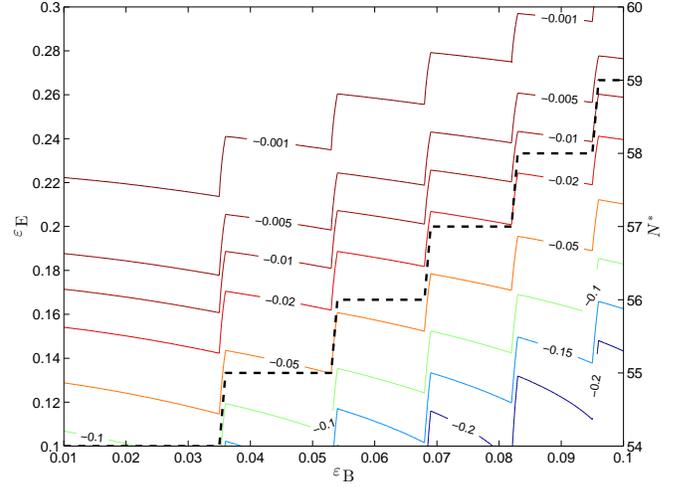}
\caption{Contour map (solid lines) depicting the loss in intercept probability caused by the change from UT to FT, as a function of $\varepsilon_\textrm{E}$ and $\varepsilon_\textrm{B}$. The value of $N^*$ (dashed line) as a function of $\varepsilon_\textrm{B}$ has been superimposed on the plot.}
%\vspace{-5mm}
\label{fig:secrecy_gain}
\end{figure}
% --- END of FIG3 ---

% -----------------------------------------------------------------------------

\section{Numerical and Analytical Results}
\label{sec:results}

This section compares the derived analytical expressions with simulation results, establishes their validity and obtains solutions to the RAM problem for various channel conditions.

Fig.~\ref{fig:fig_compare} depicts the relationship between the intercept probability and the quality of Bob's and Eve's channels, represented by $\varepsilon_\textrm{B}$ and $\varepsilon_\textrm{E}$, respectively. For each point, the value of the $N$ coded packet transmissions was optimized by RAM for $K\!=\!50$ source packets, $\hat{N}\!=\!150$ maximum allowable coded packet transmissions, a field size of $q\!=\!2$ and a target probability of Bob recovering the source message equal to $\hat{P}=90\%$. In simulations, Alice broadcasts the optimal number of coded packets determined by RAM. Instances where Eve successfully recovers $K$ linearly independent coded packets are counted and averaged over $10^4$ realizations to obtain the intercept probability. We observe the close agreement between analytical and simulation results, which confirms the tightness of \eqref{eq:prob_ic_FT} and \eqref{eq:prob_ic_UT}. Fig.~\ref{fig:fig_compare} also shows that when the channel quality between Alice and Eve is significantly worse than the channel quality between Alice and Bob, the intercept probability is close to zero for both FT and UT. As expected, the intercept probability increases when the two channels experience identical or relatively similar conditions but FT offers a clear advantage over UT. For example, for $\varepsilon_\textrm{B}=0.09$ and $\varepsilon_\textrm{E}=0.1$, the intercept probability will reduce from $68\%$ to $45\%$ if the mode of operation switches from UT to FT. The reduction in the intercept probability due to the adoption of FT becomes pronounced when $\varepsilon_\textrm{E}$ drops below $0.25$.
  
Fig.~\ref{fig:secrecy_gain} quantifies the loss in intercept probability or, equivalently, the gain in secrecy that occurs by changing the operational mode from UT to FT, as noted in \eqref{eq:diff}. The optimum value of $N$, denoted by $N^*$, has also been plotted in Fig.~\ref{fig:secrecy_gain} (secondary $y$-axis on the right-hand side of the plot). Observe that as $\varepsilon_\textrm{B}$ increases from $0.01$ to $0.1$, Alice increases the coded packet transmissions from 54 to 59 in an effort to maintain the probability of Bob recovering the source message at \mbox{$\hat{P}=90\%$}. Notice the abrupt change in the intercept probability each time RAM generates a new optimum value for $N$, based on $\varepsilon_\textrm{B}$.

A way to reduce the intercept probability, especially in settings where the values of $\varepsilon_\textrm{B}$ and $\varepsilon_\textrm{E}$ are similar, has been hinted in the Remark. If Alice can measure the instantaneous quality of the channel between her and Bob and transmits coded packets only when the measured quality is above an acceptable threshold, the effective value of $\varepsilon_\textrm{B}$ will be reduced and the intercept probability will drop at the expense of delay.

% -----------------------------------------------------------------------------

% -- MOVE Fig.2 HERE for camera-ready

\section{Conclusion}
\label{sec:conclusion}

We derived accurate expressions for the intercept probability of a network, where a transmitter uses random linear network coding to broadcast information. Both unaided transmission and feedback-aided transmission were investigated and the secrecy gain achieved by the latter approach was computed. We presented a resource allocation model to minimize the intercept probability, while satisfying delay and reliability constraints, and showed that the legitimate receiver is not required to have knowledge of the presence of an eavesdropper. Theoretical and simulation results identified the channel erasure probabilities for which feedback-aided transmission offers a lower intercept probability than unaided transmission when the proposed resource allocation model is applied.

% -----------------------------------------------------------------------------

\appendix

% -- Appendix 1 --

\subsection{Reformulation of the intercept probability of FT}
\label{sec:rewrite_FT}

Based on the definition of the PMF in \eqref{eq:prob_pmf}, the expression for $P_{\,\textrm{BE}}(N)$ in \eqref{eq:prob_BE} can be expanded as follows:
\begin{equation}
\label{eq:prob_BE_step1}
\begin{split}
P_{\,\textrm{BE}}(N)=&\;F_\textrm{B}(K)F_\textrm{E}(K)\\
&-F_\textrm{B}(K)F_\textrm{E}(K+1)+F_\textrm{B}(K+1)F_\textrm{E}(K+1)\\
&-\dots\\
&-F_\textrm{B}(N-1)F_\textrm{E}(N)+F_\textrm{B}(N)F_\textrm{E}(N).
\nonumber
\end{split}
\end{equation}
If we create pairs from each two consecutive terms, with the exception of the last term, and invoke again the definition of the PMF, we obtain
\begin{equation}
\label{eq:prob_BE_step2}
P_{\,\textrm{BE}}(N)=\left[\;-\!\!\!\!\sum_{n_\textrm{T}=K+1}^{N}\!\!f_\textrm{E}(n_\textrm{T})\,F_\textrm{B}(n_\textrm{T}-1)\right]+F_\textrm{B}(N)F_\textrm{E}(N).
\nonumber
\end{equation}
In \eqref{eq:prob_E}, we established  that $P_{\,\textrm{E}}(N)=F_\textrm{E}(N)-F_\textrm{B}(N)F_\textrm{E}(N)$. Using \eqref{eq:prob_ic_FT}, the intercept probability of FT can be expressed as:  
\begin{equation}
\label{eq:prob_ic_FT_new}
P^{\,\textrm{FT}}_\textrm{int}(N) = F_\textrm{E}(N)-\!\!\sum_{n_\textrm{T}=K+1}^{N}\!\!f_\textrm{E}(n_\textrm{T})\,F_\textrm{B}(n_\textrm{T}-1).
\end{equation}

% -- Appendix 2 --

\subsection{Proof of Proposition \ref{pro:prop1} for the case of FT}
\label{sec:proofProp}
In order to prove Proposition \ref{pro:prop1} for the FT mode, it suffices to set $\Delta=P_\textrm{int}(N_2)-P_\textrm{int}(N_1)$ and show that $\Delta\geq0$ for all $N_2\geq N_1$. Using \eqref{eq:prob_ic_FT_new}, we find that
\begin{equation}
\label{eq:proof_step1}
%\begin{split}
\Delta=\,F_\textrm{E}(N_2)-F_\textrm{E}(N_1)-\!\!\!\!\sum_{n_\textrm{T}=N_1+1}^{N_2}\!\!\!f_\textrm{E}(n_\textrm{T})\,F_\textrm{B}(n_\textrm{T}-1).
%\end{split}
\end{equation}
Terms $-F_\textrm{E}(i)$ and $F_\textrm{E}(i)$ for $i=N_1+1,\ldots,N_2-1$, which cancel each other out, are added to $F_\textrm{E}(N_2)-F_\textrm{E}(N_1)$ and give
\begin{equation}
\label{eq:proof_step2}
\begin{split}
F_\textrm{E}(N_2)-F_\textrm{E}(N_1)&=\,\left(F_\textrm{E}(N_2)-F_\textrm{E}(N_2-1)\right)+\ldots\\
&\quad\;\ldots+\left(F_\textrm{E}(N_1+1)-F_\textrm{E}(N_1)\right)\\
&=\!\!\!\sum_{n_\textrm{T}=N_1+1}^{N_2}\!\!\!f_\textrm{E}(n_\textrm{T}).
\end{split}
\end{equation}
If we substitute \eqref{eq:proof_step2} into \eqref{eq:proof_step1}, we obtain
\begin{equation}
\label{eq:proof_step3}
\Delta=\!\!\!\!\sum_{n_\textrm{T}=N_1+1}^{N_2}\!\!\!\!f_\textrm{E}(n_\textrm{T})\bigl[1-F_\textrm{B}(n_\textrm{T}-1)\bigr]\nonumber
\end{equation}which is a sum of non-negative terms and is, thus, $\Delta\geq 0$.

% -- MOVE Fig.3 HERE for camera-ready

% -----------------------------------------------------------------------------
\bibliographystyle{IEEEtran}
\bibliography{IEEEabrv,CL2015_1842_Ref}
% -----------------------------------------------------------------------------

\end{document}